\documentclass[a4paper,11pt]{article}
\usepackage{pos,ulem}
\pdfoutput=1
\makeatletter
     {\bgroup\raggedright\small\section*{\refname
        \@mkboth{\MakeUppercase\refname}{\MakeUppercase\refname}}%
      \list{\name{bib\@arabic\c@enumiv}
            \@biblabel{\@arabic\c@enumiv}}%
           {\settowidth\labelwidth{\@biblabel{#1}}%
            \setlength\itemsep{0pt}
            \setlength\parskip{0pt}
            \setlength\parsep{0pt}
            \setlength\partopsep{0pt}
            \leftmargin\labelwidth
            \advance\leftmargin\labelsep
            \@openbib@code
            \usecounter{enumiv}%
            \let\p@enumiv\@empty
            \renewcommand\theenumiv{\@arabic\c@enumiv}}%
      \sloppy\clubpenalty4000\widowpenalty4000%
      \sfcode`\.\@m}
     {\def\@noitemerr
       {\@latex@warning{Empty `thebibliography' environment}}%
      \endlist\egroup}
\makeatother

\newcommand{\qsqmin}[0]{q^2_\text{min}}
\newcommand{\qsqmax}[0]{q^2_\text{max}}
\newcommand{\w}[0]{\omega}

\title{Nonperturbative calculations of form factors for exclusive semileptonic $B_{(s)}$ decays}

\author[a]{Jonathan M.~Flynn}
\author[a,b]{Ryan C.~Hill}
\author[a]{Andreas J\"uttner}
\author[c]{Amarjit Soni}
\author[d]{\newline J.~Tobias Tsang}
\author*[e,f]{Oliver Witzel}

\affiliation[a]{Physics and Astronomy, University of Southampton,
  Southampton SO17 1BJ, UK}
\affiliation[b]{DISCnet Centre for Doctoral Training, University
  of Southampton, Southampton SO17 1BJ, UK}
\affiliation[c]{Physics Department, Brookhaven National Laboratory,
  Upton, NY 11973, USA}
\affiliation[d]{CP3-Origins and IMADA, University of Southern Denmark,
  Campusvej 55, DK-5230 Odense M, Denmark}
\affiliation[e]{Department of Physics, University of Colorado, Boulder, CO 80309, United States}
\affiliation[f]{Theoretische Physik 1, Naturwissenschaftlich-Technische Fakultät,
Universität Siegen, 57068 Siegen, Germany}

\emailAdd{r.c.hill@soton.ac.uk}
\emailAdd{oliver.witzel@uni-siegen.de}

\abstract{Precise theoretical predictions derived from the Standard
  Model are a key ingredient in searches for new physics in the flavor
  sector. The large mass and long lifetime of the $b$ quark make
  processes involving $b$ quarks of particular interest. We use
  lattice simulations to perform nonperturbative QCD calculations for
  semileptonic $B_{(s)}$ decays. We present results from our
  determinations of $B_s\to D_s \ell \nu$ and $B_s\to K \ell \nu$
  semileptonic form factors and provide an outlook for our $B\to
  \pi\ell\nu$ calculation. In addition we discuss the determination of
  $R$-ratios testing lepton-flavor universality and suggest use of an
  improved ratio. Our calculations are based on the set of 2+1 flavor
  domain wall Iwasaki gauge field configurations generated by the
  RBC-UKQCD collaboration featuring three lattice spacings of $1/a =
  1.78$, $2.38$, and $2.79\,\text{GeV}$. Heavy $b$-quarks are
  simulated using the relativistic heavy quark action.\\[3mm]

\textbf{For the RBC and UKQCD collaborations}}

\FullConference{%
  40th International Conference on High Energy physics - ICHEP2020\\
  July 28 - August 6, 2020\\
  Prague, Czech Republic (virtual meeting)
}

\begin{document}
\maketitle

\section*{Introduction}
Semileptonic decays of $B_{(s)}$ mesons play an important role in
testing and constraining the Standard Model (SM) of elementary
particle physics. Theoretical calculations of such processes, in
combination with experimental data, allow extraction of
Cabibbo-Kobayashi-Maskawa (CKM) matrix elements, or comparison of
theory and experiment for ratios testing lepton flavor universality.
Focusing on exclusive semileptonic decays, we use lattice quantum
chromodynamics (QCD) simulations to obtain from first principles the
form factors parametrizing the contributions due to the strong force.
Specifically we report on our work for $B_s\to D_s\ell\nu$ and $B_s\to
K\ell\nu$ decays and in addition present an outlook on our calculation
of $B\to\pi\ell\nu$ decays. Each of these processes can be described
by two form factors, $f_+$ and $f_0$, which parametrize the
semileptonic decay rate
\begin{multline}
\label{eq:B_semileptonic_rate}
  \frac{d\Gamma(B_{(s)}{\to} P\ell\nu)}{dq^2} =
  \eta_\text{EW} \frac{G_F^2 |V_{xb}|^2}{24\pi^3} \,
  \frac{(q^2{-}m_\ell^2)^2 |\vec k|}{(q^2)^2}
    \bigg[ \Big(1{+}\frac{m_\ell^2}{2q^2}\Big)
           \vec k^{\,2}|f_+(q^2)|^2\\
      +\frac{3m_\ell^2}{8q^2}
      \frac{(M_{B_{(s)}}^2{-}M_P^2)^2}{M_{B_{(s)}}^2}|f_0(q^2)|^2
    \bigg]\,,
\end{multline}
where $x=c$ for $P=D_s$ and $x=u$ for $P=K,\,\pi$. The corresponding
meson masses are denoted by $M_{B_{(s)}}$ and $M_P$ and the
$4$-momenta by $p$ and $k$, respectively. $E$ labels the energy of the
$P$-meson and $q = p-k$ is the momentum transfer with $|\vec k| =
(E^2-M_P^2)^{1/2}$. $G_F$ is the Fermi constant and $\eta_\text{EW}$
an electroweak correction factor. After presenting our form factors
for the three different processes, we discuss testing lepton flavor
universality by defining $R$-ratios of form factors integrated over
different ranges in $q^2$.

Our numerical work is based on using RBC-UKQCD's set of 2+1 flavor
gauge field configurations
\cite{Allton:2008pn,Aoki:2010dy,Blum:2014tka,Boyle:2017jwu} featuring
the physical effects of two degenerate up/down quarks as well as the
strange quark in the sea sector and inverse lattice spacings of 1.78,
2.38, and $2.79\,\text{GeV}$. Light, strange, and charm quarks are simulated
using a domain-wall action
\cite{Kaplan:1992bt,Shamir:1993zy,Furman:1994ky,Blum:1996jf,Blum:1997mz,Brower:2012vk},
whereas bottom quarks are simulated using the relativistic heavy quark
action \cite{Christ:2006us,Lin:2006ur,Aoki:2012xaa}. In total we use
six different ensembles with pion masses down to $267\,\text{MeV}$.

\section*{Semileptonic \texorpdfstring{$B_s\to D_s\ell\nu$ and $B_s\to K\ell\nu$}{Bs->Ds l nu and Bs -> K l nu} decays}
The LHCb collaboration recently presented the first experimental
results for exclusive semileptonic $B_s$ decays
\cite{Aaij:2020hsi,Aaij:2020xjy}. With improved statistical precision
and a finer resolution of the $q^2$ bins, $B_s$ decays will provide an
interesting, alternative channel to determine the CKM matrix elements
$|V_{cb}|$ and $|V_{ub}|$ which so far are extracted using $B\to
D^{(*)}\ell\nu$ and $B\to\pi\ell\nu$ decays, respectively. Eventually
$B_s$ decays may help shed light on the persistent discrepancy between
inclusive and exclusive determinations~\cite{Gambino:2019sif} or help
address the tension with the SM observed in the ratio $R_{D^{(*)}}$
testing lepton flavor universality. The nonperturbative lattice
calculation favors $B_s$ decays over $B$ decays because both numerical
costs and statistical uncertainties grow with decreasing light-quark
mass. Calculations for $B_s$ mesons are therefore computationally
cheaper with smaller statistical errors than comparable calculations
for $B$ mesons. In addition the larger mass of the hadronic final
state in the $B_s$ decay reduces the amount of energy released, which
makes the calculation slightly easier.

\begin{figure}[t]
  \includegraphics[height=0.168\textheight]{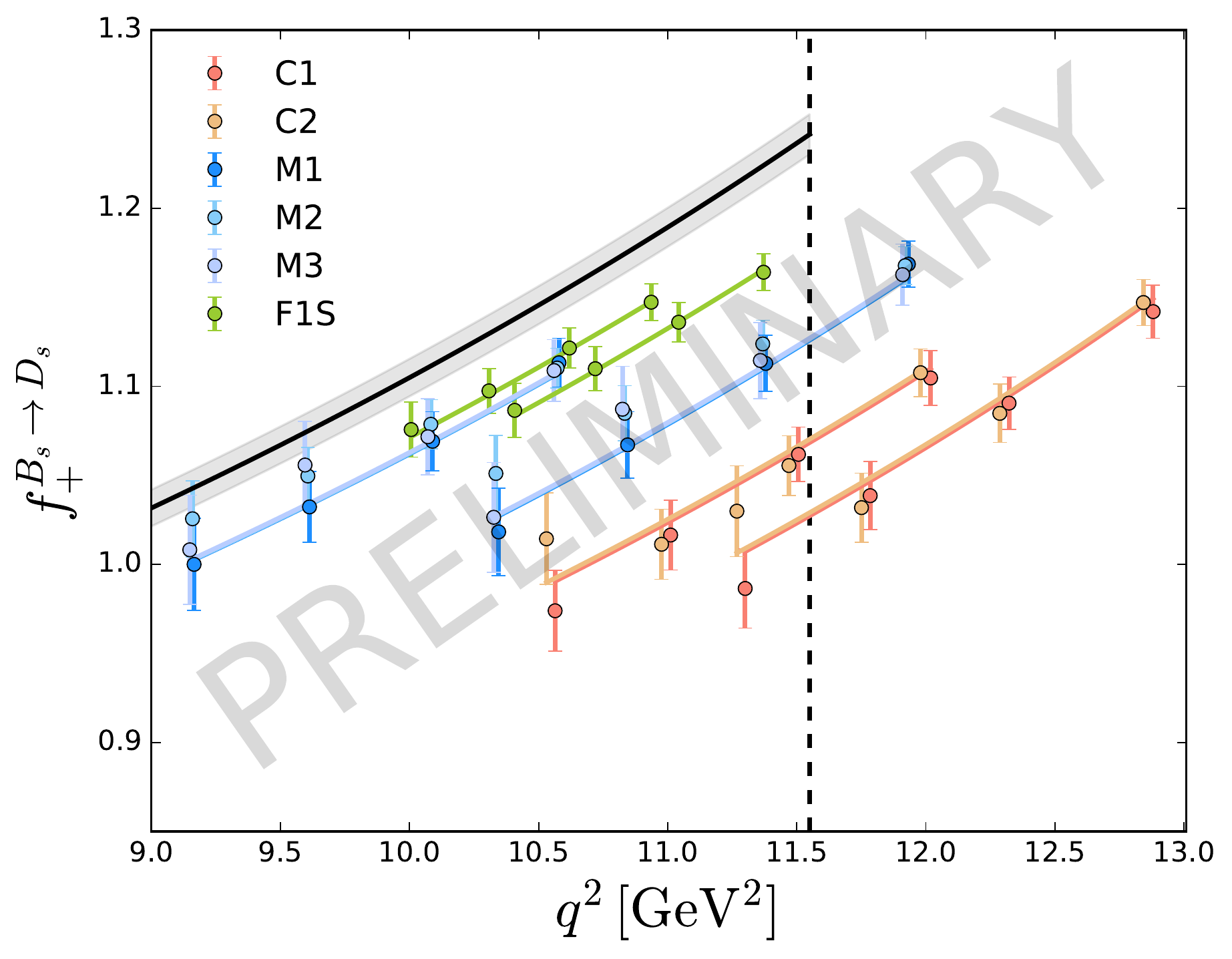}
  \includegraphics[height=0.168\textheight]{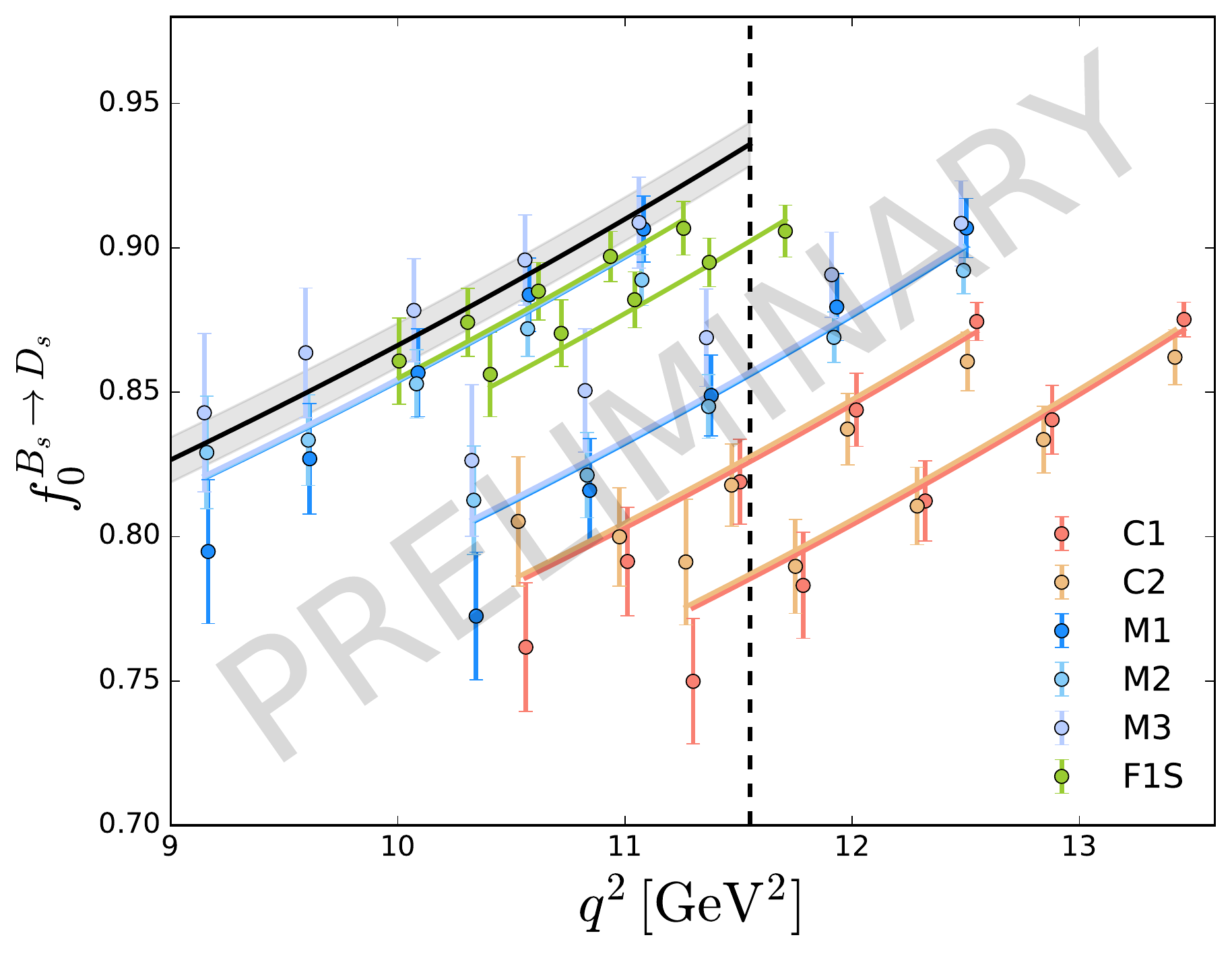}
  \includegraphics[height=0.168\textheight]{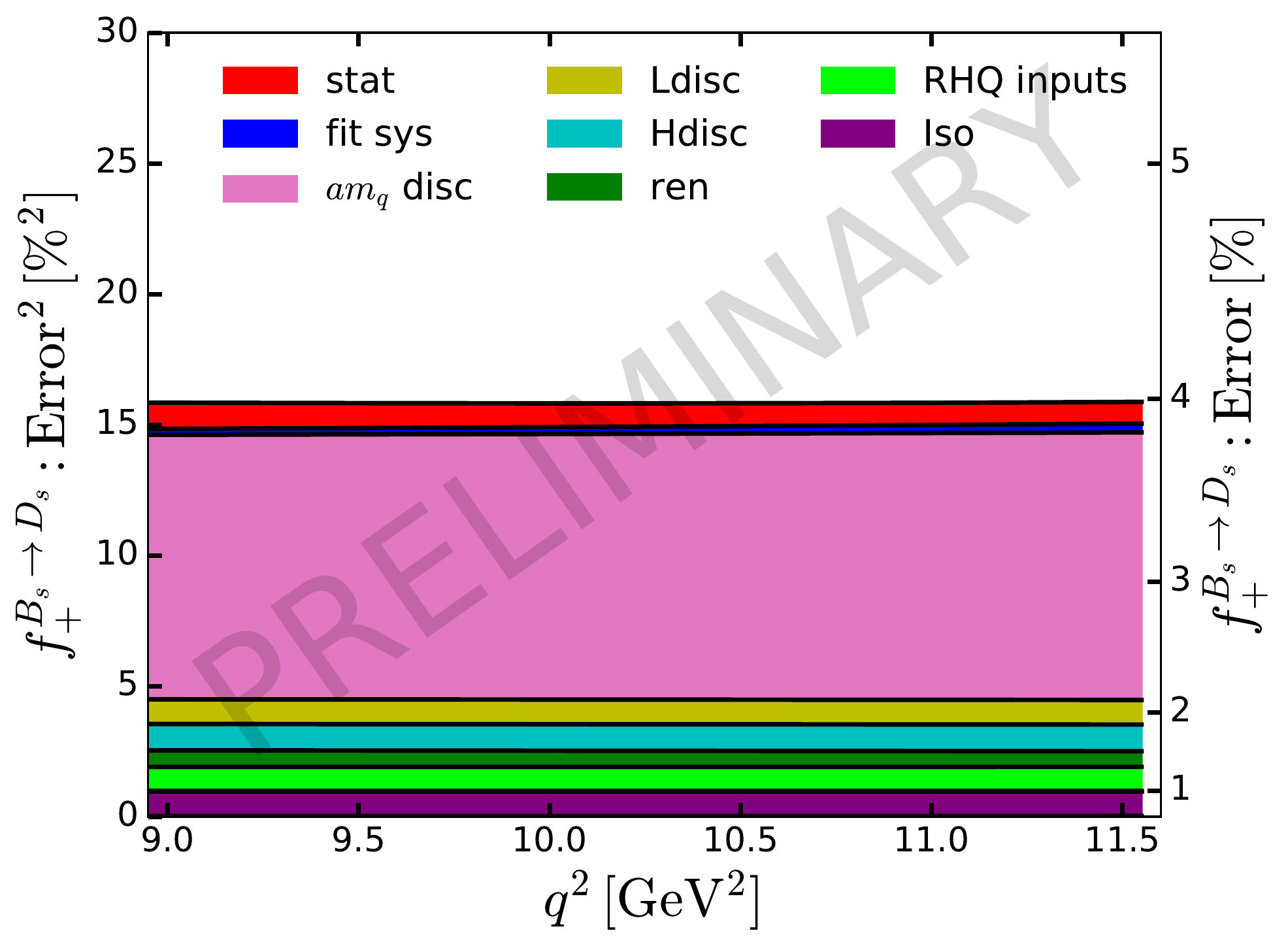}
\caption{Form factors $f_+$ and $f_0$ for semileptonic $B_s\to
  D_s\ell\nu$ decays. The left and central panels show the outcome of
  our numerical simulations as well as the chiral-continuum limit
  obtained from a combined and correlated fit to all data points. The
  right panel presents our full error budget for $f_+$ which is
  presently dominated by the uncertainty due to discretizing the charm quarks.}
\label{fig.BsDs}
\end{figure}

We present our results for the $f_+$ and $f_0$ form factors for
$B_s\to D_s\ell\nu$ decays in the left and central panels of
Fig.~\ref{fig.BsDs}. The colored data points are the outcome of our
numerical simulations on the six different gauge field ensembles and
the color indicates the lattice spacing: in orange coarse ensembles
(C), in blue medium ensembles (M), and in green the fine ensemble (F).
Next we perform a combined and correlated fit of all data points to
obtain the chiral-continuum limit (black line with gray error band),
i.e.~the form factor at physical quark masses in the continuum for the
range of $q^2$ we directly access in our simulation. This global fit
is based on a polynomial ansatz using a low order Pad\'e
approximation. The right panel gives a preview of our full error
budget where in addition to the statistical uncertainty (red), we
consider variations to the fit function (blue) and estimate further
systematic effects. At present the uncertainty due to the
discretization of the charm quark (magenta) dominates the error
budget. While we show the error budget only for $f_+$, that for $f_0$
looks similar.

\begin{figure}[t]
  \includegraphics[height=0.168\textheight]{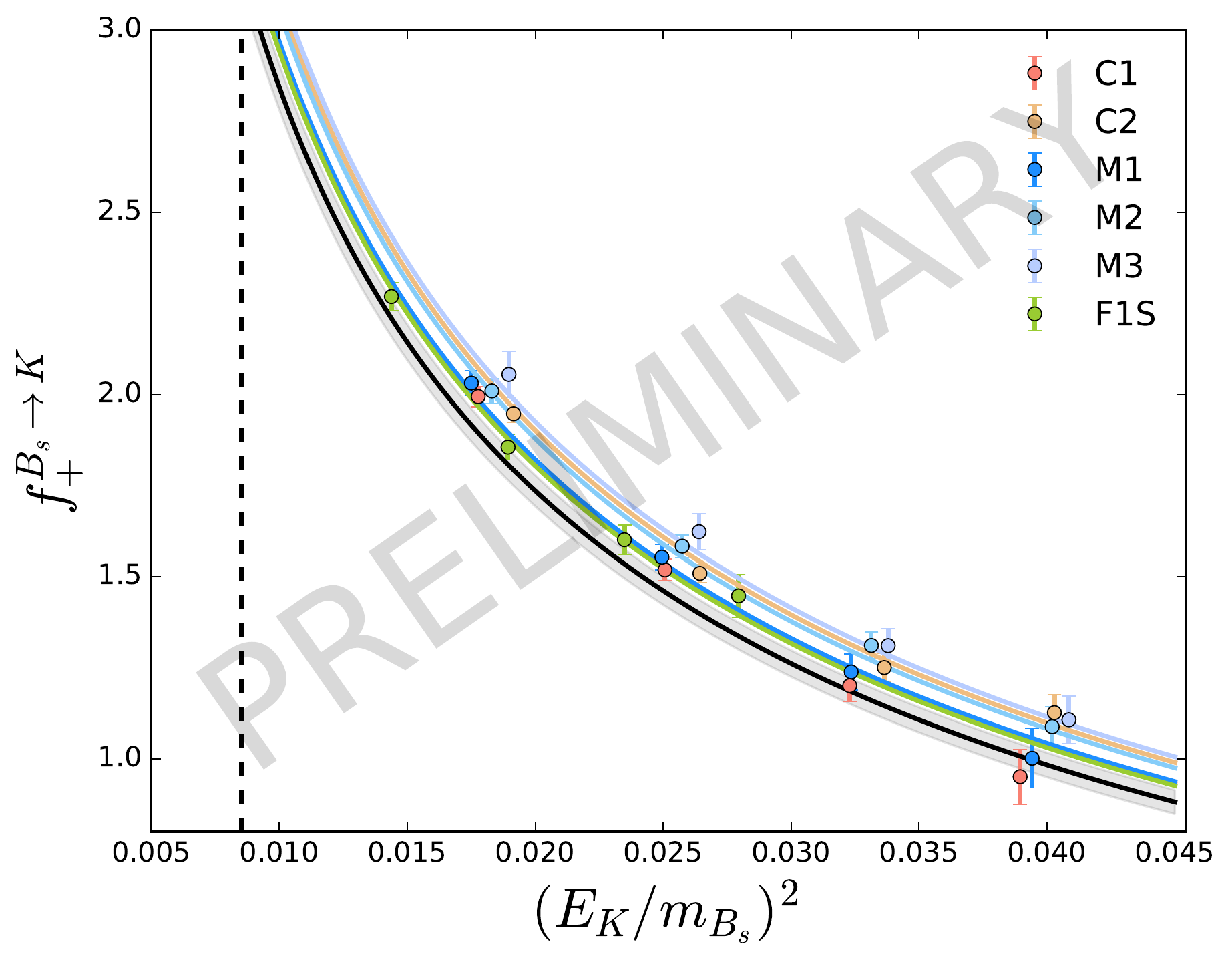}
  \includegraphics[height=0.168\textheight]{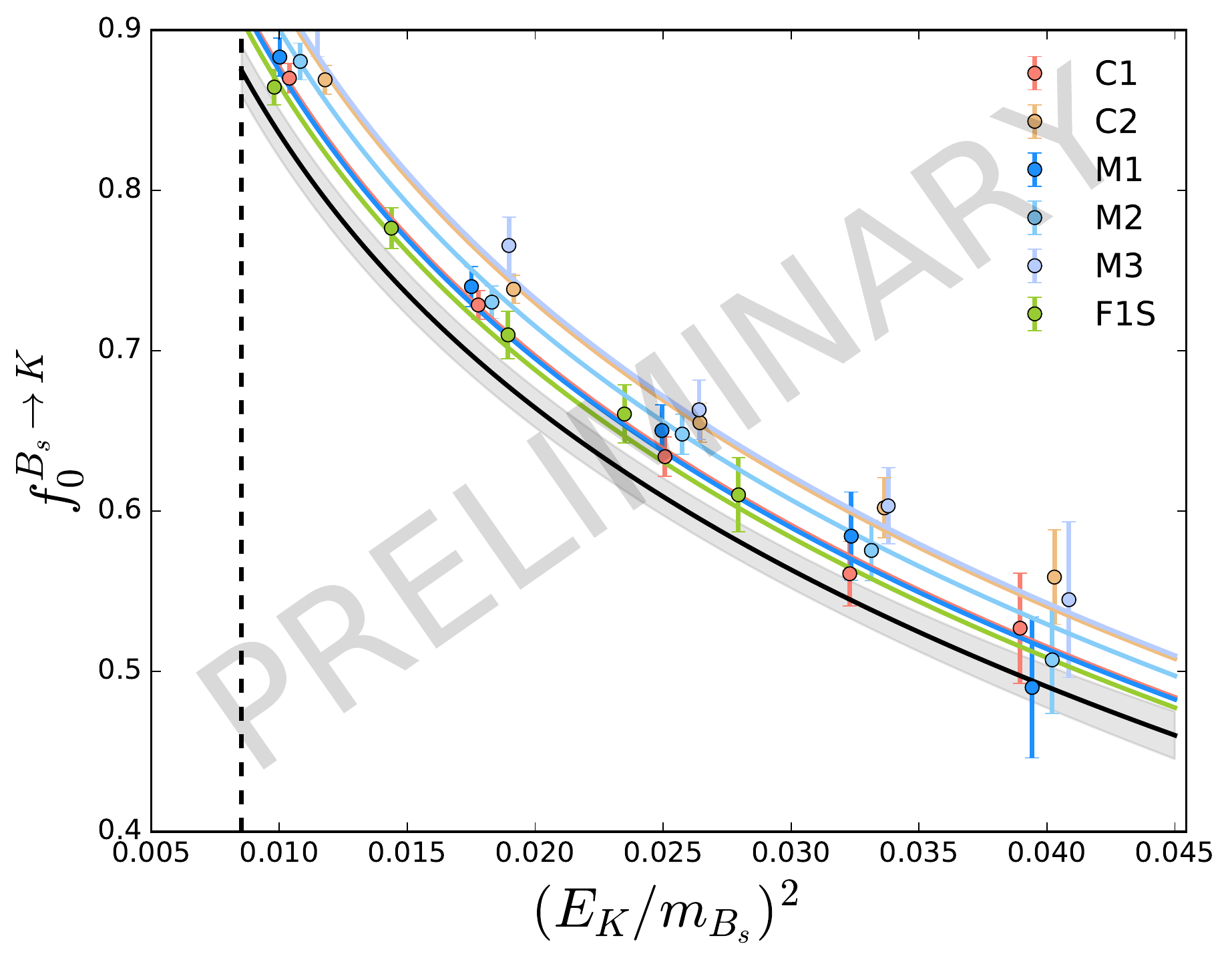}  
  \includegraphics[height=0.168\textheight]{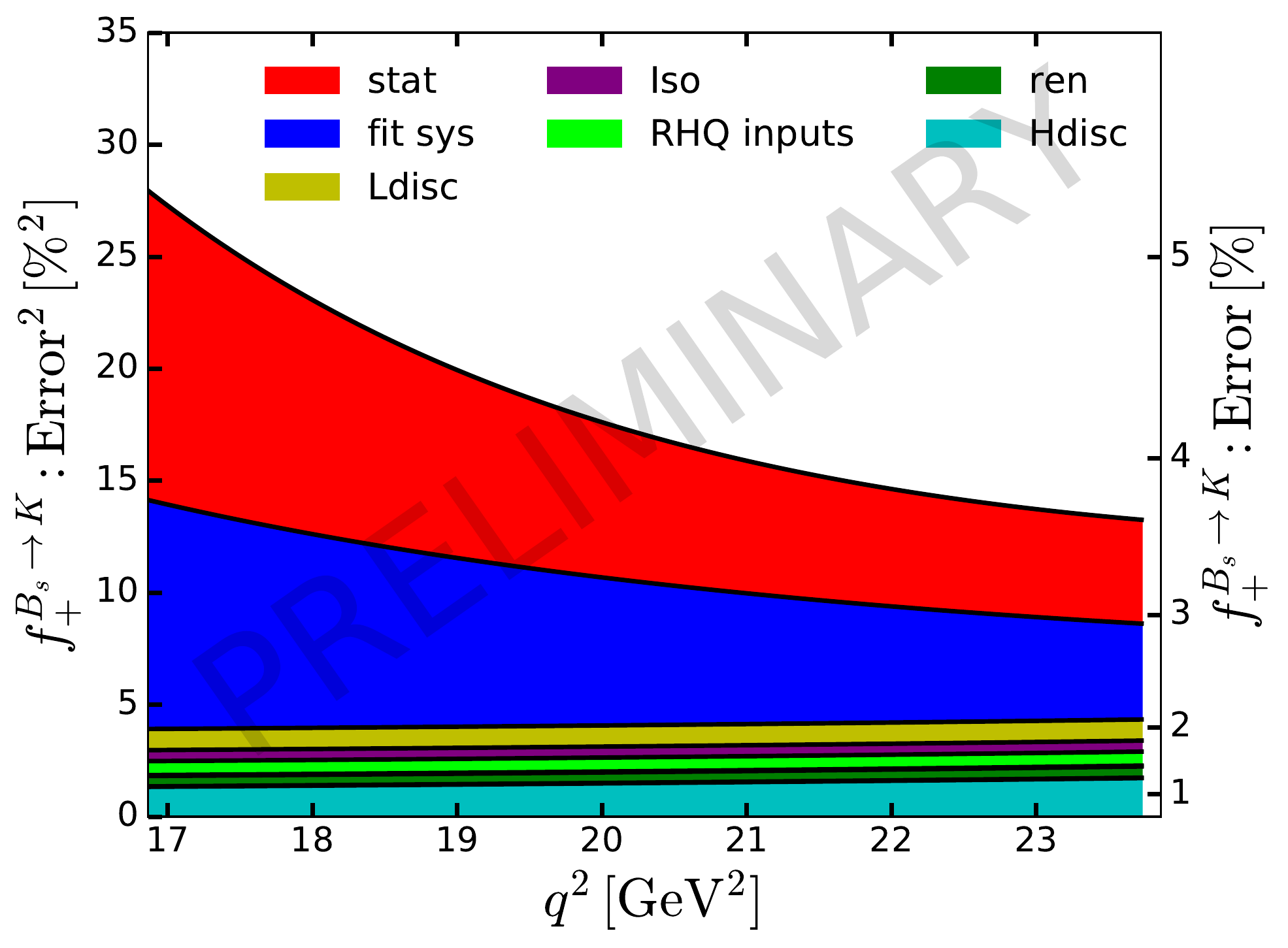}
\caption{Form factors $f_+$ and $f_0$ for semileptonic $B_s\to K\ell\nu$ decays (left and central panel) and full error budget for $f_+$ (right panel) where at present effects due to statistical uncertainties dominate.}
\label{fig.BsK}
\end{figure}

In Fig.~\ref{fig.BsK} we similarly show the status of our analysis for
$B_s\to K\ell\nu$. The left and central panels again show the form
factors as we calculate them on the lattice in combination with the
result in the chiral-continuum limit. For $B_s\to K\ell\nu$ we obtain
the latter by performing a global fit to all data points using an
ansatz inspired by SU(2) heavy meson chiral perturbation theory
(HM$\chi$PT) in the limit for hard kaons. As the panel on the right
indicates, our results are at present dominated by the statistical
uncertainties.

With the full results of lattice calculation for $B_s\to D_s\ell\nu$
and $B_s\to K\ell\nu$ form factors to hand, we are currently studying
the kinematical extrapolation of our form factors over the full range
of $q^2$ using the $z$ parametrization following Boyd, Grinstein and
Lebed (BGL)~\cite{Boyd:1994tt} or Bourrely, Caprini and Lellouch
(BCL)~\cite{Caprini:1997mu}. Scrutinizing the $z$ expansion for
possible systematic effects, for example due to the order of the
expansion, is important in order to obtain reliable predictions for
ratios testing lepton flavor universality since these are computed by
integrating the decay rate over physically allowed regions in $q^2$.

\section*{Semileptonic \texorpdfstring{$B\to\pi\ell\nu$}{B -> pi l nu} decays}
\begin{figure}[t]
  \includegraphics[height=0.16\textheight]{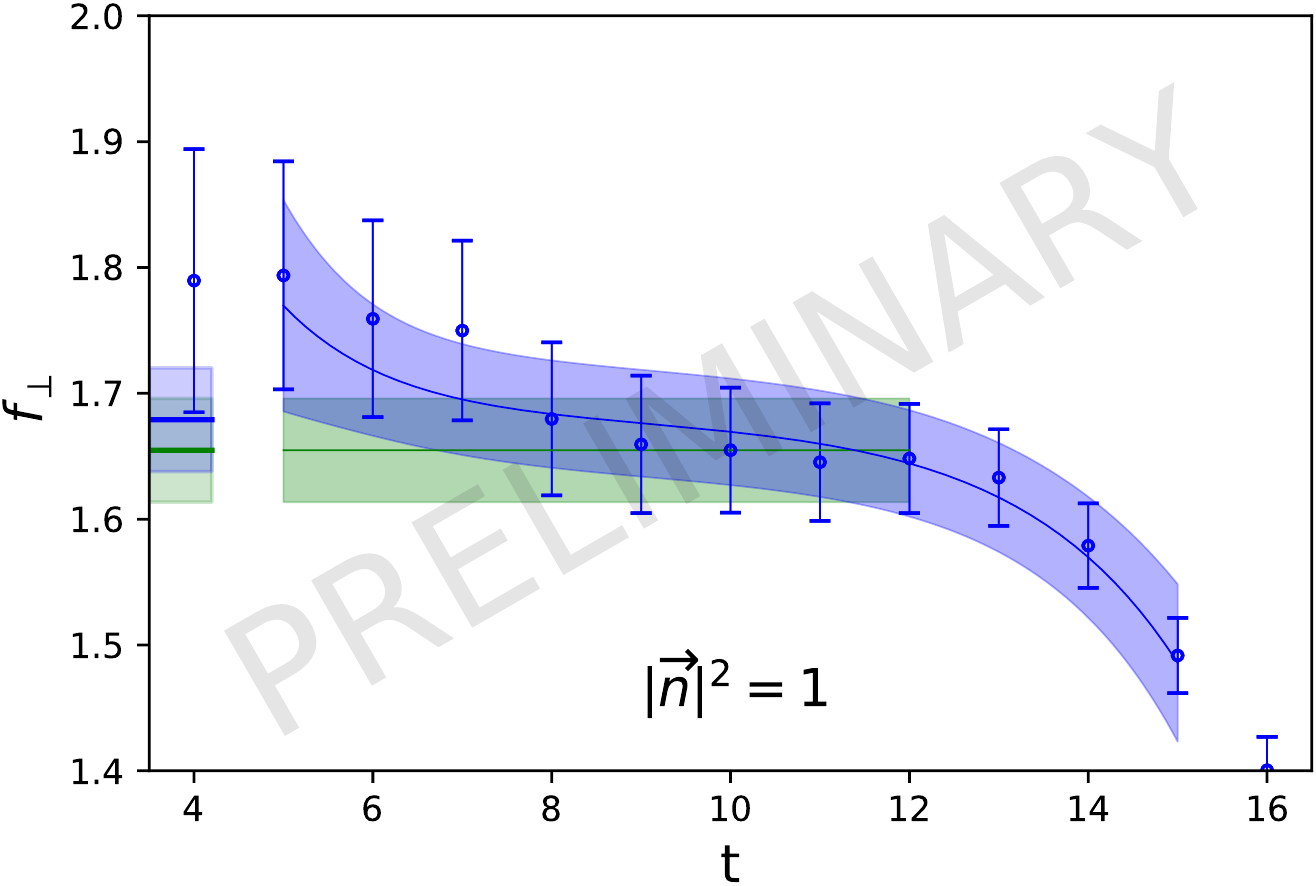}
  \includegraphics[height=0.16\textheight]{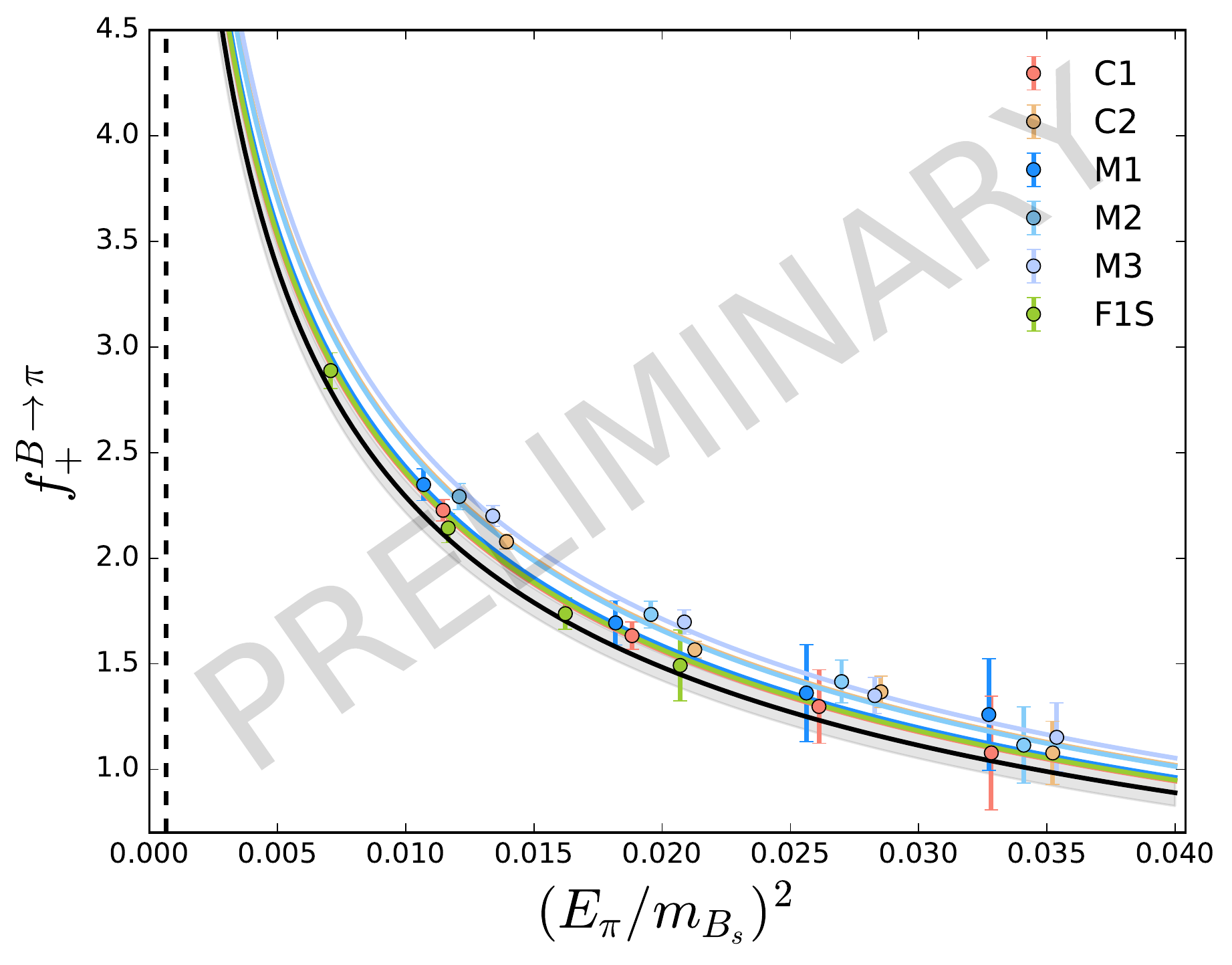}
  \includegraphics[height=0.16\textheight]{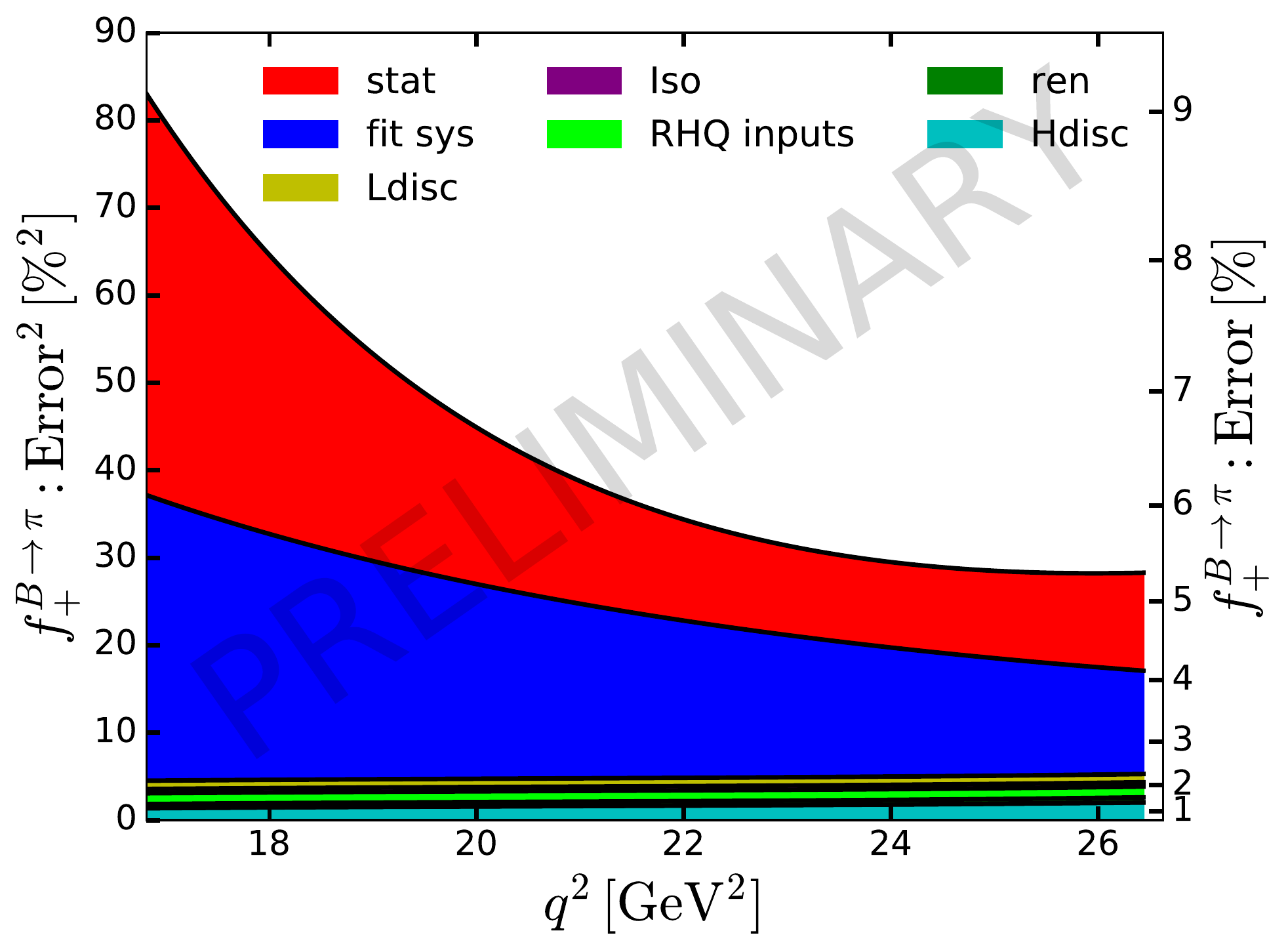}    
\caption{View of our $B\to\pi\ell\nu$ calculation. Left: comparison
  between ground state and excited state fit to extract the form
  factor signal with one unit of momentum on the coarse C1 ensemble.
  Center: Lattice determination of the form factor $f_+$ as well as
  the chiral-continuum limit obtained from an SU(2) HM$\chi$PT fit.
  Right: preliminary error budget for $f_+$.}
\label{fig.Bpi}
\end{figure}

In parallel we are working on an update of our $B\to\pi\ell\nu$
calculation from 2015~\cite{Flynn:2015mha}. Including the new fine
ensemble (F1S) at a third, finer lattice spacing combined with
improved estimates for systematic effects and an enhanced analysis, we
intend to obtain an improved result with reduced uncertainties. In the
first step of the analysis, we refine the extraction of the form
factors by including additional terms parametrizing one excited state
for both the $B$ meson as well as the pion. This leads to a larger
range of time slices entering the determination and hence a better
estimate as can be seen in the left most plot of Fig.~\ref{fig.Bpi}
where a simple groundstate fit (green) is compared to the one
including excited states (blue). Extracting values for the form factors on
all six ensembles and using up to four units of momenta, we show in
the central panel our values to determine $f_+$ combined with the
chiral-continuum limit obtained using SU(2) HM$\chi$PT. A preview of
the corresponding full error budget is shown in the right most panel of
Fig.~\ref{fig.Bpi} where again effects due to statistical
uncertainties dominate.

\section*{Ratios testing lepton flavor universality violations}

Lepton flavor universality is an accidental symmetry in the SM. One
way to test it is to consider semileptonic decays with different
leptonic ($e$, $\mu$, $\tau$) final states. Although $e$, $\mu$ and
$\tau$ have the same couplings to $W$ and $Z$ bosons in the SM, their
different masses mean that the shapes of differential decay widths
with respect to $q^2$ differ, as do (partially) integrated decay
rates. A commonly-performed test of the SM is to compare experimentally
measured and theoretically predicted values for the ratio
  \begin{equation}
    R_{D_s^{(*)}}= 
      \int_{m_\tau^2}^{q^2_\text{max}}dq^2\,
      \frac{d\Gamma(B_s\to D_s^{(*)}\tau\bar\nu_\tau)}{dq^2}
      \Bigg/
      \int_{m_l^2}^{q^2_\text{max}}dq^2\,
      \frac{d\Gamma(B_s \to D_s^{(*)}l\bar\nu_l)}{dq^2},
  \label{Eq.RDs}
  \end{equation}
where the integration is taken from $q^2 = m_{l(\tau)}^2$ to the
maximum value of $q^2$ that is kinematically allowed, and $l$ denotes
either $e$ or $\mu$.

This definition of $R_{D_s^{(*)}}$ has the drawback that the
denominator contribution from $m_l^2 \leq q^2 \leq m_\tau^2$ has no
correspondence in the numerator and hence cannot provide useful
information to test lepton flavor universality. We therefore
follow~\cite{Atwood:1991ka,Isidori:2020eyd} and consider a different ratio
\begin{equation}
  R^\text{imp}_{D_s^{(*)}} =
  \int_{\qsqmin}^{\qsqmax} dq^2\,
  \frac{d\Gamma(B_s\to D_s^{(*)}\tau\bar\nu_\tau)}{dq^2}
  \Bigg/
  \int_{\qsqmin}^{\qsqmax} dq^2\,
     \left[\frac{\w_\tau(q^2)}{\w_l(q^2)}\right]\,
   \frac{d\Gamma(B_s\to D_s^{(*)}l\bar\nu_l)}{dq^2},
  \label{Eq.RDs_UN}
\end{equation}
using a common integration range in the numerator and
denominator with lower limit $\qsqmin \geq m_\tau^2$, and
reweighting the integrand in the denominator with the factor
$\w_\tau(q^2)/\w_l(q^2)$, where
\begin{equation}
  \w_\ell(q^2) = \bigg(1-\frac{m_\ell^2}{q^2}\bigg)^2
               \bigg(1+\frac{m_\ell^2}{2q^2}\bigg)\quad
               \text{for}\quad
               \ell=e,\mu,\tau.         
\end{equation}
This form of the ratio was introduced in~\cite{Isidori:2020eyd} for
hadronic vector final states and we adopt it for pseudoscalar final
states such as for $B_s\to D_s\ell\nu$ decays. The contribution of the scalar
form factor in the denominator of Eq.~(\ref{Eq.RDs_UN}) can be neglected because
$m_\mu^2/(m_\mu^2+2q^2)  \ll 1$ for $q^2 \in [m_\tau^2,\qsqmax)$ 
and we use the decay rate expression from
Eq.~(\ref{eq:B_semileptonic_rate}) to find
\begin{align}
  R^\text{imp,SM}_{D_s} &= 1 + \frac{\int_{\qsqmin}^{\qsqmax} dq^2\,
   \Phi(q^2) \w_\tau(q^2) (F_S^\tau)^2}
  {\int_{\qsqmin}^{\qsqmax} dq^2\,
    \Phi(q^2) \w_\tau(q^2) F_V^2},
  \intertext{with $\Phi(q^2)=\eta_\text{EW}G_F^2 |V_{xb}|^2\vec
    k/24\pi^3$ and}
  (F_S^\tau)^2 &= \frac{3}{4} \frac{m_\tau^2}{m_\tau^2+2q^2}
  \frac{(M^2-m^2_\tau)^2}{M^2}\, |f_0(q^2)|^2  \label{eq:FSlcontrib}  \\
  F_V^2 &= \vec k^2 |f_+(q^2)|^2.
\end{align}

\section*{Acknowledgments}\vspace{-2mm}
We thank our RBC and UKQCD collaborators for helpful discussions and
suggestions. Computations used resources provided by the USQCD
Collaboration, funded by the Office of Science of the US~Department of
Energy and by the \href{http://www.archer.ac.uk}{ARCHER} UK National
Supercomputing Service, as well as computers at Columbia University
and Brookhaven National Laboratory. We used gauge field configurations
generated on the DiRAC Blue Gene~Q system at the University of
Edinburgh, part of the DiRAC Facility, funded by BIS National
E-infrastructure grant ST/K000411/1 and STFC grants ST/H008845/1,
ST/K005804/1 and ST/K005790/1. This project has received funding from
Marie Sk{\l}odowska-Curie grant 659322 (EU Horizon 2020), STFC grants
ST/P000711/1 and ST/T000775/1. RH was supported by the DISCnet Centre
for Doctoral Training (STFC grant ST/P006760/1). OW acknowledges
support from DOE grant DE--SC0010005. AS was supported in part by US
DOE contract DE--SC0012704. JTT acknowledges suppport from the
Independent Research Fund Denmark, Research Project~1, grant
8021--00122. No new experimental data was generated. The project
leading to this application has received funding from the European
Union’s Horizon 2020 research and innovation programme under the Marie
Sk{\l}odowska-Curie grant agreement No 894103.

{\small
  \bibliography{../B_meson}
  \bibliographystyle{JHEP-notitle}
}


\end{document}